\documentstyle[epsfig]{mn}
\begin{document} 
\LARGE
\normalsize
 
\title[Radio emission in Wd1]
{Discovery of extended radio emission in the young cluster Wd1}

\author[J.S. Clark et al]{J.S. Clark$^{1,2}$, R.P. Fender$^{1,3}$,
L.B.F.M. Waters$^3$, S.M.  Dougherty$^4$, J. Koornneef$^5$,
\and
 I. A. Steele$^6$, and A. van Blokland$^3$\\
$^1$ Astronomy Centre, University of Sussex, Falmer, Brighton, BN1
9QH, UK\\
$^2$ Department of Physics and Astronomy, University of Southampton,
Southampton, SO17 1BJ, UK \\
$^3$ Astronomical Institute Anton Pannekoek, University of Amsterdam
and Center for High Energy Astrophysics, Kruislaan 403,\\ 1098 SJ
Amsterdam, Netherlands \\
$^4$ University of Calgary/DRAO, PO Box 248, White Lake Rd, Penticon, B.C.,
V2A 6K3 Canada \\
$^5$ Kapteyn Astronomical Institute, University of Groningen,
Groningen, 9700 AV, The Netherlands\\
$^6$ Astrophysics Group, Liverpool John Moores University, Liverpool,
L3 3AF, UK \\
}
\date{Version 21 May 1998}

\maketitle
\begin{abstract}

We present 10 $\mu$m, ISO-SWS  and Australia Telescope Compact
Array observations of the region in the cluster Wd1 in Ara centred on
the B[e] star Ara C. An ISO-SWS spectrum reveals
 emission from highly ionised species
in the vicinity of the star, suggesting a secondary source of
excitation in the region. We find strong radio emission at both
3.5cm and 6.3cm, with a total spatial extent of over 20 arcsec.  The
emission is found to be concentrated in two discrete structures,
separated by $\sim$14''. The westerly source is resolved,
with a spectral index indicative of thermal emission.  
The easterly source
is clearly extended and nonthermal (synchrotron) in
nature. Positionally, the B[e] star is found to coincide with the more
compact radio source, while the southerly lobe of the extended source
is coincident with Ara A, an M2 I star.  Observation of
the region at 10$\mu$m reveals strong emission with an almost
identical spatial distribution to the radio emission.
Ara C is found to have an extreme radio luminosity in comparison
to prior radio observations of hot stars such as O and B
supergiants and Wolf-Rayet stars, given the estimated distance to the
cluster. 
An origin in a detatched shell of material around the central
star is therefore suggested; however given the spatial extent of the emission,
such a shell must be relatively young ($\tau$$\sim$10$^3$yrs).
The extended non thermal
emission associated with the M star Ara A is unexpected; to the best
of our knowledge this is a unique phenomenon. SAX (2-10keV) observations
show no evidence of X-ray emission, which might be expected if a
compact companion were present.

\end{abstract}
\begin{keywords}
stars:individual(Ara C) - stars:emission line, B[e] - stars:radio emission
\end{keywords}

\section{Introduction}

B[e] stars are characterised by a large infrared excess, thought
to be due to hot dust at $\sim$10$^3$K. In this respect the B[e] stars
differ from classical Be stars, where the IR excess is
due to free free emission from hot gas. Spectroscopically B[e] stars
differ from classical Be stars due to the presence of narrow forbidden
line emission from singly ionised metals, such as Fe$^+$ in
addition to strong Balmer line emission at optical wavelengths. UV
spectra reveal species with a much higher degree of ionisation than
seen in optical spectra, in an
analogous manner to classical Be stars. This fact led Zickgraf et
al. (1986) to propose a two component wind to
explain the spectra, with a slowly expanding cool equatorial component
responsible for the optical spectral features, and a hotter, high
velocity polar component producing the high excitation features in the
UV spectra.
Near IR observations of  B[e] stars by McGregor et
al. (1988) and subsequently  Morris et al. (1996) reveal 
Na\, {\sc i} and CO bandhead emission, confirming the presence of
cooler circumstellar regions as indicated by the dust emission. 

The spectra also show a  similarity
between B[e] stars and both Be supergiant and LBV spectra. On the
basis of the morphological similarities between the spectra they
suggest a close evolutionary link between the three types of star. If
confirmed this would be an important result, as there is still some
confusion as to the exact evolutionary status of B[e] stars (see for
example Lamers et al. 1998). In order to investigate the
evolutionary status of these objects, observations of the galactic B[e]
star Ara C were made as it is the
only known galactic B[e] star in a cluster, permitting an 
accurate measurement of its luminosity, and hence age (since 
very luminous young B[e] stars are not expected to be seen).

\section{The Ara cluster}

The diffuse object in Ara found at 2.2$\mu$m by Price (1968) was
identified by Westerlund (1968) as a heavily reddened cluster. Further
work by Borgman et al. (1970), Koornneef (1977) and most recently
Westerlund (1987) have done much to clarify the properties of the
cluster. The cluster is thought to be young, with an age of only
$\sim$7 million years (Westerlund 1987). The distance is estimated  to
be $\sim$5.2kpc, which gives a cluster size of some 3pc, within which
photometry is available for
some $\sim$258 stars (Westerlund 1987). The cluster is heavily
reddened, with $A_{\rm V} \sim 11$, of which $A_{\rm V}=7$  is thought to
be due to dust in the region of the cluster. Westerlund (1987)
estimates a minimum mass of dust of 
$\sim$20M$_{\odot}$, which implies the presence of $\sim$2000M$_{\odot}$
of gas compared to 1200M$_{\odot}$ to be found in 
82 of the brightest cluster stars
(lower limiting luminosity of log {\em L}/{\em L}$_{\odot}$=3.5). The
diffuse H$\alpha$ emission seen in images of the cluster is attributed to
excitation by the stellar radiation fields.

The intrinsically brightest stars of the cluster form a continuous
sequence from B2 Ia to M2 Ia. Below the supergiant sequence Westerlund
(1987) finds that most of the stars are luminosity class II-III, with
the faintest stars on the main sequence. Ara C, identified as star 9
in Westerlund (1987), was of immediate interest due to 
a remarkable emission line spectrum, consisting of very strong Balmer 
emission (EW$_{H{\alpha}}{\sim}$250$\AA$), HeI emission, 
as well as FeII, [NII], [OII] and 
[ArIII]. Westerlund classified the star as Be, but the
H$\alpha$ equivalent 
width is at least a factor 5 higher than the largest values reported
for classical Be stars, and comparison with the spectra of 
other massive emission line objects suggests that star C 
may be a B[e] star. Due to uncertainty in the intrinsic colours of
B[e] stars, the values of $M_v \sim$6 and $A_v \sim$8.70 derived for
Ara C in Westerlund (1987) are the subject of some
uncertainty. Likewise, the estimate of log({\em
L}$_{\ast}$/{\em L}$_{\odot})$=4.86 for Ara C was derived under the
assumption of T$_{eff} \sim$20000K, and should be treated with caution.

Many B[e] stars are known in our Galaxy, but their nature is unclear,
since their spectral characteristics coincide closely with those of 
massive YSO's. The situation would improve if B[e] stars could be
identified in clusters, as is proposed for Ara C. 
This implies that Ara C is a post-main-sequence object
since the age of the cluster is such that massive stars can no longer be 
YSO's. This is, to the best of our 
knowledge, the first case in which a galactic 
B[e] star can definitively be classified as a post-main-sequence object, 
which demonstrates that  Galactic counterparts to the LMC B[e] 
supergiants, as discussed by Zickgraf et al. (1986), do exist. 

\section{Observations}

\subsection{ISOCAM, ISO-SWS and TIMMI observations}

ISOCAM images at $\sim$3-10$\mu$m of the Ara C region were made on
1996 September 10. The ISOCAM data were taken with the
1.5 pixel field of view (FOV) setting. 
The array is 32x32 pixels and so the total FOV
is about 47x47 arcsec (the PSF for the wavelengths used can be
calculated from Cesarsky et al 1996). The 3$\mu$m image shows 
Ara C to be an unresolved point source (several other bright stars in
the cluster are also visible as point sources). Ara C is also visible
in the 10$\mu$m image as a point source, as is  Ara A (nomenclature
by Borgman et al 1970; denoted star {\em 26} by Westerlund 1987). 

Based on a
combination of low resolution optical spectroscopy and photometry
Westerlund (1987) classifies this as an M2 I star, with a bolometric
luminosity of log({\em L}$_{\ast}$/{\em L}$_{\odot})$=5.26 and an
effective temperature of log({\em T}$_{eff}$)=3.54. Further
extended emission is visible to the north of this source in the
10$\mu$m image. A 10$\mu$m TIMMI image is shown in Figure 2, and confirms
the the identification of extended emission to the north of Ara A 
(TIMMI is fully described in 
K\"aufl et al 1992; the broad band N filter was used in the
observations). Due
to its absence from the 3$\mu$m image this is most likely due to
emission from cold dust (compared to
the  warmer dust thought to be responsible for the unresolved emission
centred on Ara C and Ara A).

A fast SWS spectrum centred on Ara C was also obtained at this time
(ISO-SWS is fully described in De Graauw et
al. 1996). The spectral range employed was
2.4-45$\mu$m; gaussian fits to the strongest forbidden lines are
presented in Figure 3.
This was expected to yield information on the
presence and composition of circumstellar dust, and forbidden line
emission (we note that based on the presence of both emission lines
in the spectrum, and hot dust associated with the star,
 the classification of Ara C as a B[e] star is confirmed).
There is no evidence in the spectrum for the presence of the Unidentified 
Infrared (UIR) emission bands 3.29, 6.2, 7.6/7.7, 8.6, 11.3, and 12.7$\mu$m 
that tend to occur together. These features are referred to by Tokunaga 
(1996) as `carbonaceous emission bands', and are sometimes attributed to 
polycyclic aromatic hydrocarbons (PAHs), which might be expected to
arise in the dusty circumstellar envelope. 
However, a  strong narrow
emission feature is visible at 20.59$\mu$m, possibly
due to a
simple oxide (although an association with crystaline silicates cannot
be ruled out at this time). The spectrum also 
revealed the presence of  [O\, {\sc iv}]. These observations are deeply
puzzling, as this observation  implies temperatures of $\sim$80,000K, which
cannot be due to the stellar radiation field ($\sim$20,000K), and
clearly imply the presence of another source of ionising radiation in
the system. It should be noted that there is likely to be some
contamination from the M2 I star Ara A in the spectrum, and so the
exact location of the highly ionised gas cannot be conclusively
identified with the circumstellar environment of Ara C from this
observation alone.

\subsection{Radio Observations}

In a survey of bright Southern radio sources Wright et al. (1994) report a 
detection  of a point source 
at co-ordinates RA 16 47 07.5, Dec -45 50 39.0 
(J2000; designated PMNJ 1647-4550 in their catalogue), 
$\sim$20'' of RA from the optical co-ordinates reported for Ara A.
They estimate an error of 6.9'' in RA, making 
the position  of the point source  $\sim$3$\sigma$ from the extended near-IR 
source reported here. However, no other radio source within 5 arcmin is
reported by the SIMBAD database, 
and so we conclude that despite the discrepancy in 
the positions, PMNJ 1647-4550 is likely to correspond to the resolved emission 
associated with Ara A and C.

Observations of the Ara C region at 3.6 and 6.3cm  were obtained with
the Australian Telescope Compact Array (ATCA) on 1997 April 30. These
observations were interleaved with those of a nearby  phase referenced
source (1646-50), with a total on source integration of 93
minutes. The flux scale was determined by observations of
J1934-638. After determining initial antennae gain solutions by phase
referencing, the gain solutions were improved by three iterations of
self-calibration. 

Strong radio emission at both 3.5 and 6.3 cm was detected at the
expected position of Ara C, and in an extended structure $\sim 10$
arcsec to the east (Figure 1). The eastern emission is concentrated into
two lobes which we label Ara A (North) and Ara A (South) respectively.
We find that the emission associated with Ara C is
thermal in nature (positive spectral index) while that of Ara A(N \& S) 
is nonthermal in origin (negative spectral index).
Examination of the radio images suggests that Ara C can be closely
represented by a 2D gaussian. To estimate the size of Ara C we have
examined the visibility data directly. 
After locating the peak of Ara C at the phase-tracking centre of the
image, we removed the flux contributed from Ara A(S) and Ara A(N)
using the {\sc AIPS} routine {\sc UVSUB}. The remaining visibilities
were then binned into a $15\times15$ grid to improve the
signal-to-noise ratio. 
These visibilities clearly show that Ara C is resolved by the array.
A two-dimensional Gaussian fit to this gridded
data then gives an estimate of the source size, the total flux of the
source and the position angle of the emission distribution 
(see Tables 1 and 2). For comparison, we also
estimated the source size by fitting two-dimensional-Gaussian
functions to the image of Ara C ({\sc AIPS} routine {\sc
JMFIT}). These give the same results as the visibility fitting within
the uncertainties. 

\begin{figure}
\leavevmode\epsfig{file=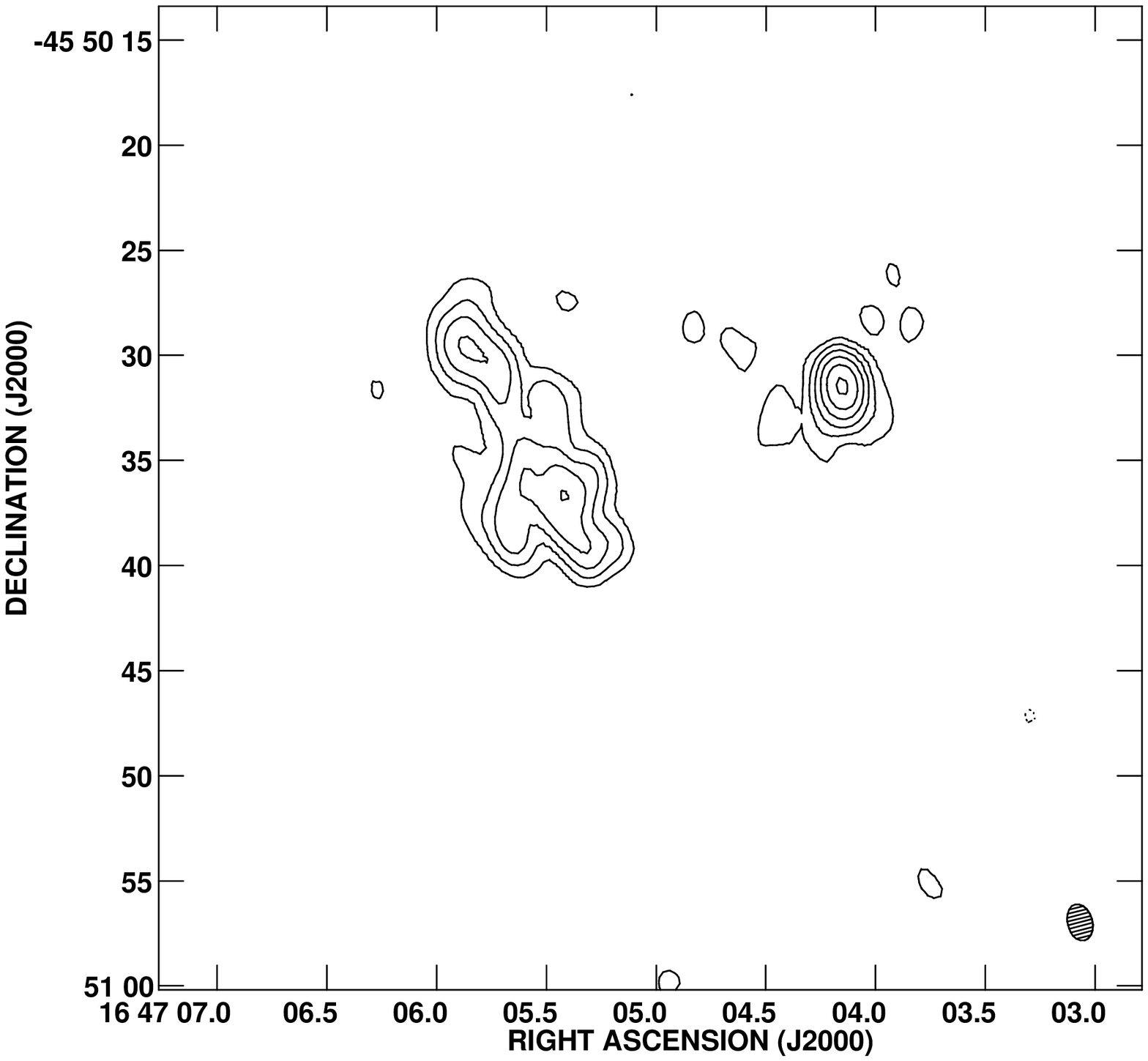,width=7cm,clip=,angle=0,bblly=159,
bbllx=-20,bbury=670,bburx=570}
\leavevmode\epsfig{file=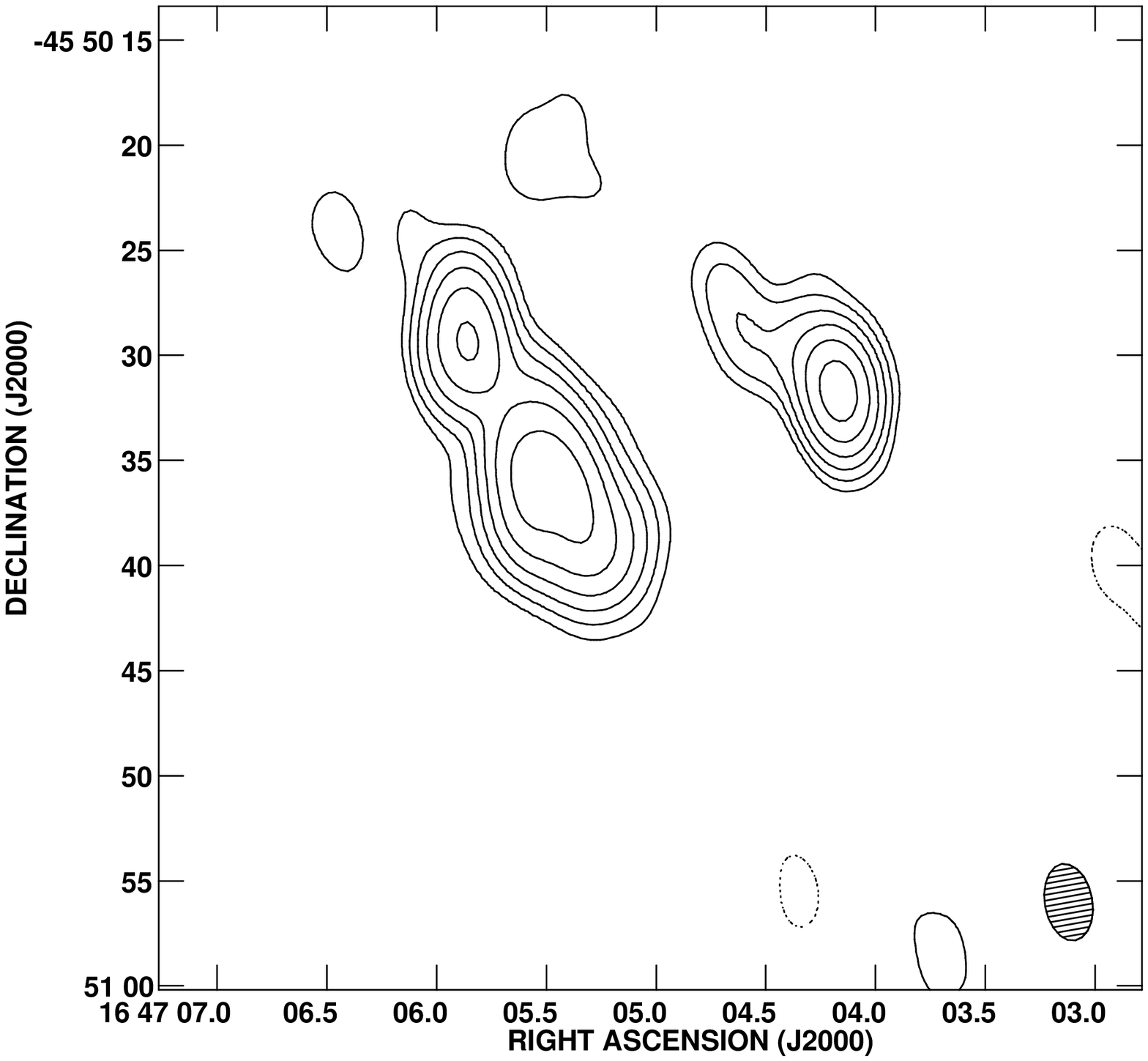,width=7cm,clip=,angle=0,bblly=159,
bbllx=-20,bbury=670,bburx=570}

\caption{
3.5 cm  (upper) and 6.3 cm (lower) maps of the region surrounding 
Ara A and Ara C (contours are at -1, 1, 2, 4, 8, 16, 32 ,64, 128 and
256 times the r.m.s. noise level of 1.1mJy and 1.5mJy
respectively). Beam sizes are indicated in the lower right corner of
each panel (1.8$\times$1.2 arcsecs at 14 degrees and
3.7$\times$2.3 arcsecs at 10 degrees for 3.5cm and 6.3cm respectively).
} 
\end{figure}

\begin{figure}
\leavevmode\epsfig{file=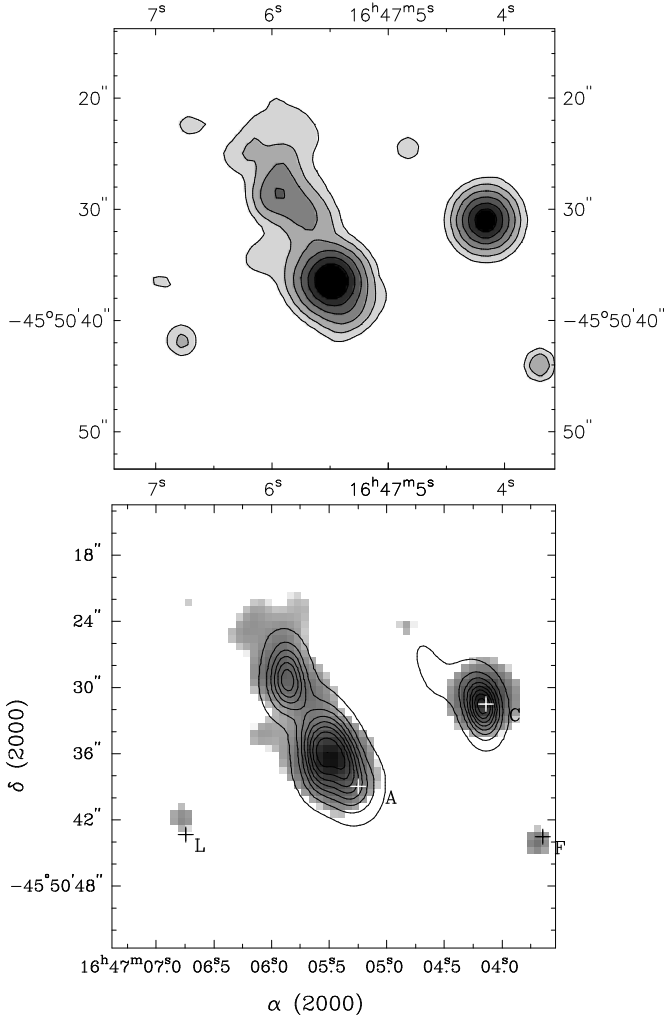,width=6.8cm,clip=,angle=270,bblly=67,
bbllx=191,bbury=574,bburx=574}
\vspace*{3mm}
\leavevmode\epsfig{file=overlay.ps,width=6.8cm,clip=,angle=270,bblly=58,
bbllx=107,bbury=519,bburx=522}

\caption{Upper Panel:
TIMMI 10$\mu$m image of the field (fluxes derived from ISOCAM
observations of $\sim$50Jy and $\sim$190Jy  for Ara C and Ara A respectively). 
Lower Panel: overlay of the TIMMI (greyscale) and
6.3cm (contour) plots. The positions of other cluster stars
(notation by Borgman et al 1970) are also shown.}
\end{figure}

The coordinates of
the Gaussian fit to the western compact radio condensation are RA 16
47 04.2 Dec -45 50 31.6 (J2000). An estimate for the position of Ara C
of RA 16 47 04.2 Dec -45 50 30.7 was
derived from a combination of the ISO-CAM image and an STScI DSS image
(ie within 1 arcsec of the radio co-ordinates).
In Figure 2 we overplot the 10$\mu$m TIMMI image on the 6.3cm map,
centering both image and map on Ara C (ISOCAM fluxes for Ara C and Ara A of
$\sim$50Jy and $\sim$190Jy respectively). We find an excellent
coincidence between emission at both wavelengths in both lobes of  Ara
A. Optical co-ords for Ara A(S) are RA 16 47 05.25 Dec -45 50 39.  
Radio co-ords for Ara A(S) and Ara A(N) are RA 16 47 05.42 Dec -45 50 36.7 and 
RA 16 47 05.85 Dec -45 50 29.66 respectively. 
We further plot the positions of other cluster stars on
this image  (notation by Borgman et al 1970). We cite the excellent
agreement of these positions with the TIMMI image as evidence for the
identification of Ara C with the westerly radio source. We further note that
our co-ordinates are consistent, within the quoted  errors, to all
prior published positions.

\begin{table}
\begin{center}
\begin{tabular}{cccc} \hline
              & \multicolumn{2}{c}{Integrated flux} & Spectral  \\
              & 3.5cm  (mJy) & 6.3cm (mJy) & Index   \\
\hline
Ara C         & $51.7 \pm 0.6$ & $41.8 \pm 0.5$ & $0.36 \pm 0.03$\\
Ara A (S) & $92.4 \pm 1.0$ & $125.4 \pm 0.8$ & $-0.52 \pm 0.02$\\
Ara A (N) & $30.1 \pm 0.7$ & $35.8 \pm 0.4$ & $-0.30 \pm 0.04$ \\
\hline
\end{tabular}
\caption {Flux data for the source coincident with 
Ara C and the 2 components of the emission
coincident with Ara A. }
\end{center}
\end{table}

\begin{table}
\begin{center}
\begin{tabular}{ccc} \hline
Wavelength & Source Size (milli arcsec) & Position Angle \\
\hline
3.5cm & $ 954 \pm 24 \times 523 \pm 18$ & $166^{\circ} \pm 3 $\\
6.3cm  & $ 1179 \pm 74 \times 732 \pm 51$ & $10^{\circ} \pm 5 $\\
\hline
\end{tabular}
\caption{Source size for Ara C (derived from visibility fitting)}
\end{center}
\end{table}

\begin{figure}
\vspace{10cm}
\caption{Line profiles for the strongest forbidden lines in the
ISO-SWS spectrum. Data points from  the up and down scans are shown
separately. The nominal (solid line) and best fit (broken line)
wavelengths are also indicated.}
\end{figure}

\section{Discussion}

The spectral index of the  source Ara C is suggestive of thermal
emission, possibly from a stellar wind.
However the spectral index ($\alpha$=0.36$\pm$0.03) clearly departs
from that expected for either an isothermal, spherically symmetric
wind of constant velocity 
($\sim$0.6)  or an optically thick wind ($\alpha$=2.0). Given the
distance estimates of $\sim$5.2kpc, the implied radio luminosity is in
excess of any previously observed OB supergiant or Wolf Rayet,
comparable only to certain Luminous Blue Variables (LBVs). If the
radio luminosity is attributed solely to a stellar wind, this implies an
excessive mass loss rate given the bolometric
luminosity of the star. Likewise,
the fact that the emission is extended argues against attributing
the flux to a stellar wind; the emitting region is
two orders of magnitude larger than the effective radii for free-free radio
emission from OB stars reported by Lamers \& Leitherer 
(1993), for example. Consequently, we propose that the
radio flux is produced by a two component model consisting of a (partially)
optically thick stellar wind, and an extended optically thin 
component produced by reradiation. This could be the result of
emission from ambient intracluster material (although the previous
main sequence wind regime might have cleared a cavity around the
star), or from a detached shell of material surrounding the star.
A similar structure has been observed before in 
radio and IRAS observations of G79.29+0.46, with an optically thin
detached shell of material centred on
a stellar source with a partially optically thick 
radio spectrum (Higgs et al. 1994, Waters
et al. 1996). Waters et al. (1996) conclude that the present day wind
characteristics of this star suggest that it could be a member of the
class of Luminous Blue Variables (LBVs), while  the detatched shell is
either the result of a  Red Supergiant (RSG) phase, or an extended
period of LBV-like mass loss. 

Could Ara C also be an LBV? Based on morphological
similarities between K-band spectra of B[e] and LBVs Morris et
al. (1996) have suggested a close connection between the two types of
star. With a bolometric luminosity of log({\em L}$_{\ast}$/{\em
L}$_{\odot}$)=4.86, Ara C lies below the observed
range of luminosities for LBV's (log({\em L}$_{\ast}$/{\em
L}$_{\odot}$)$\geq$5-6) reported by  Hutsem\'{e}kers \& van Drom
(1991); however a lower luminosity limit to LBV behaviour is not known
with certainty at present. 
Such a luminosity suggests that Ara C has only recently started to
evolve off the main sequence of the HR diagram towards the RSG region,
and so doubts must exist as to whether it could have produced a  detached
shell of material in such a short period of time. We note 
that an error of  $\sim$20 per cent in the distance estimates to
the cluster, an underestimate of $A_{v}$ (possibly due to local
inhomogeneities in the intra cluster medium) or uncertainties in the
spectral type of the star (and hence bolometric correction)
could result in an  increase in the  bolometric luminosity
of Ara C to log({\em
L}$_{\ast}$/{\em L}$_{\odot}$)=5.0. Such a luminosity would then place it
in the region of the HR diagram occupied by LBVs (and would also allow
it to have evolved through a RSG phase).
Unfortunately due to lack of high resolution spectral or radio
observations  it is not possible to extract information on the present 
wind characteristics of Ara C, and so it is not possible to determine
a  present day mass loss rate or terminal wind velocity, both of which
would help to determine the evolutionary status of the star. However,
it is possible to estimate the age of the detached shell from the its
angular extent (given in Table 2). If the distance to the cluster is
taken to be 5.2kpc, and a value of 30kms$^{-1}$ is adopted for the
terminal wind velocity (appropriate for a RSG; Waters et al 1996),
then the age of the shell is found to be $\sim$600yrs. Since Maeder \&
Meynet (1988) estimate the time taken for massive stars to evolve from
the red to blue region of the HR diagram to be 1-3 10$^4$yr, this
would tend to argue against the shell being formed during a RSG phase
(although it should be noted that the higher densities of the ISM
expected in this cluster could affect the shell expansion velocity).

An explanation for the nonthermal emission positionally coincident
with Ara A is more problematic.
Previous observations of M-type giants and supergiants (for example Reid \&  
Menton 1996, Skinner et al. 1996)  have revealed them to be radio sources. 
The emission has been attributed to free-free emission in a radio photosphere
extending a few AU. However, the nonthermal nature of the emission
observed clearly eliminates this possibility. 
Radio emission from OH masers has also been observed in 
some systems, with individual masers lying at up to $\sim$500AU from the star.
However, at an 
estimated distance of $\sim$5.2kpc, the extended emission along the NE/SW axis
 is $\sim$10$^{5}$AU long, $\sim$10$^{3}$ times larger than 
the distance at which OH masers are seen. 
The low resolution optical spectrum of Westerlund (1987) shows low level 
H $\alpha$ emission, which possibly originates from  excited gas associated
with the dust seen in the TIMMI image. No other lines are seen in 
emission in this spectrum, ruling out the identification of the M2 I star as
a component in a symbiotic system, since symbiotic systems typically
show a wealth of emission lines in their spectra. 
Given the linear structure of the
emission, it is tempting to attribute it to jet emission from
accretion onto a compact object. Such a scenario
appears to be excluded by SAX 2-10keV observations, which provide an
upper limit of 10 milliCrab to any emission 
(ie a lower luminosity than expected for a
neutron star accreting at the Eddington limit; J. Heise priv. comm.),
but we note that
possible X-ray emission might be softer than that detectable by SAX,
or such a source might   be a transient source of X-rays. Likewise,
given the current observations we are unable to excluded the
possibility of a chance alignment between Ara A and  a background
radio source. 

\section{Conclusions}

Motivated by  ISO SWS observations of the source Ara C we have
obtained radio observations at 6.3cm  and 3.5cm. These observations
resulted in two strong detections, one compact source coincident  with
the optical coordinates of Ara C, and a second extended source, the
southernmost lobe of which is coincident with Ara A, an M2 I star.
An analysis of the spectral indexes of these two detections reveals
that the emission associated with Ara C is most likely thermal in
origin, while that associated with Ara A is non thermal. Comparison to
10 $\mu$m images of the region shows a remarkable coincidence between
radio and IR continuum emission (thought to arise from hot dust). 
   
The physical processes giving rise to 
 the emission are not clear. Despite the thermal nature of 
the emission associated with Ara C, it would appear unlikely that the
emission is the result of a steady stellar wind, with both luminosity
and its resolved nature arguing against such an interpretation. A more
plausible scenario would appear to be emission from a combination of a
compact, partially optically thick 
stellar wind, and an extended optically thin source, possibly a detected
shell of material formed in a possible RSG or LBV phase. 
The relationship between the eastern emission component and Ara A and
C is as yet unclear. Despite the positional coincidence between the
southernmost lobe of both radio and 10$\mu$m emission and Ara A 
it is uncertain whether the emission is
associated with this star or the result of a chance
alignment. We note that if  emission is associated with Ara A then it
would be the first observation of non thermal radio emission from an M
supergiant.

\end{document}